\documentclass[acmsmall]{acmart}
\AtBeginDocument{%
  }

\setcopyright{acmlicensed}
\copyrightyear{2026}
\acmYear{2026}
\acmDOI{XXXXXXX.XXXXXXX}
\acmConference[ETRA '26]{ACM Symposium on Eye Tracking Research \& Applications}{June 01--04, 2026}{Marrakech, Morocco}
\acmISBN{978-1-4503-XXXX-X/2018/06}

\usepackage{xcolor}
\usepackage{color}
\usepackage{verbatim}
\usepackage{multirow}

\setcopyright{none}

\begin{document}

\title[Align-to-Scale]{Align-to-Scale: Mode Switching Technique for Unimanual 3D Object Manipulation with Gaze-Hand-Object Alignment in Extended Reality}

\author{Min-yung Kim}
\email{min.kim@kaist.ac.kr}
\affiliation{%
  \institution{Graduate School of Culture Technology,}
  \institution{KAIST}
  \city{Daejeon}
  \country{Republic of Korea}}
  
\author{Jinwook Kim}
\email{jinwook.kim31@kaist.ac.kr}
\affiliation{%
  \institution{Institute of Information Electronics,}
  \institution{KAIST}
  \city{Daejeon}
  \country{Republic of Korea}}

\author{Ken Pfeuffer}
\email{ken@cs.au.dk}
\affiliation{%
 \institution{Department of Computer Science,}
  \institution{Aarhus University}
  \city{Aarhus}
  \country{Denmark}}

\author{Sang Ho Yoon}
\email{sangho@kaist.ac.kr}
\affiliation{%
  \institution{Graduate School of Culture Technology,}
  \institution{KAIST}
  \city{Daejeon}
  \country{Republic of Korea}}

\renewcommand{\shortauthors}{Kim et al.}

\begin{abstract}
    As extended reality (XR) technologies rapidly become as ubiquitous as today's mobile devices, supporting one-handed interaction becomes essential for XR. However, the prevalent Gaze + Pinch interaction model partially supports unimanual interaction, where users select, move, and rotate objects with one hand, but scaling typically requires both hands. In this work, we leverage the spatial alignment between gaze and hand as a mode switch to enable single-handed pinch-to-scale. We design and evaluate several techniques geared for one-handed scaling and assess their usability in a compound translate-scale task. Our findings show that all proposed methods effectively enable one-handed scaling, but each method offers distinct advantages and trade-offs. To this end, we derive design guidelines to support futuristic 3D interfaces with unimanual interaction. Our work helps make eye-hand 3D interaction in XR more mobile, flexible, and accessible.
\end{abstract}

\ccsdesc[500]{Human-centered computing~Interaction techniques}
\ccsdesc[500]{Human-centered computing~Mixed / augmented reality}
\ccsdesc[500]{Human-centered computing~Virtual reality}
\ccsdesc[500]{Human-centered computing~User studies}
\ccsdesc[500]{Human-centered computing~Interaction design theory, concepts and paradigms}

\keywords{Virtual Reality, Mode Switch, Scaling, Gaze, Gestures, Eye-Hand interaction}

\begin{teaserfigure}
  \includegraphics[width=\textwidth]{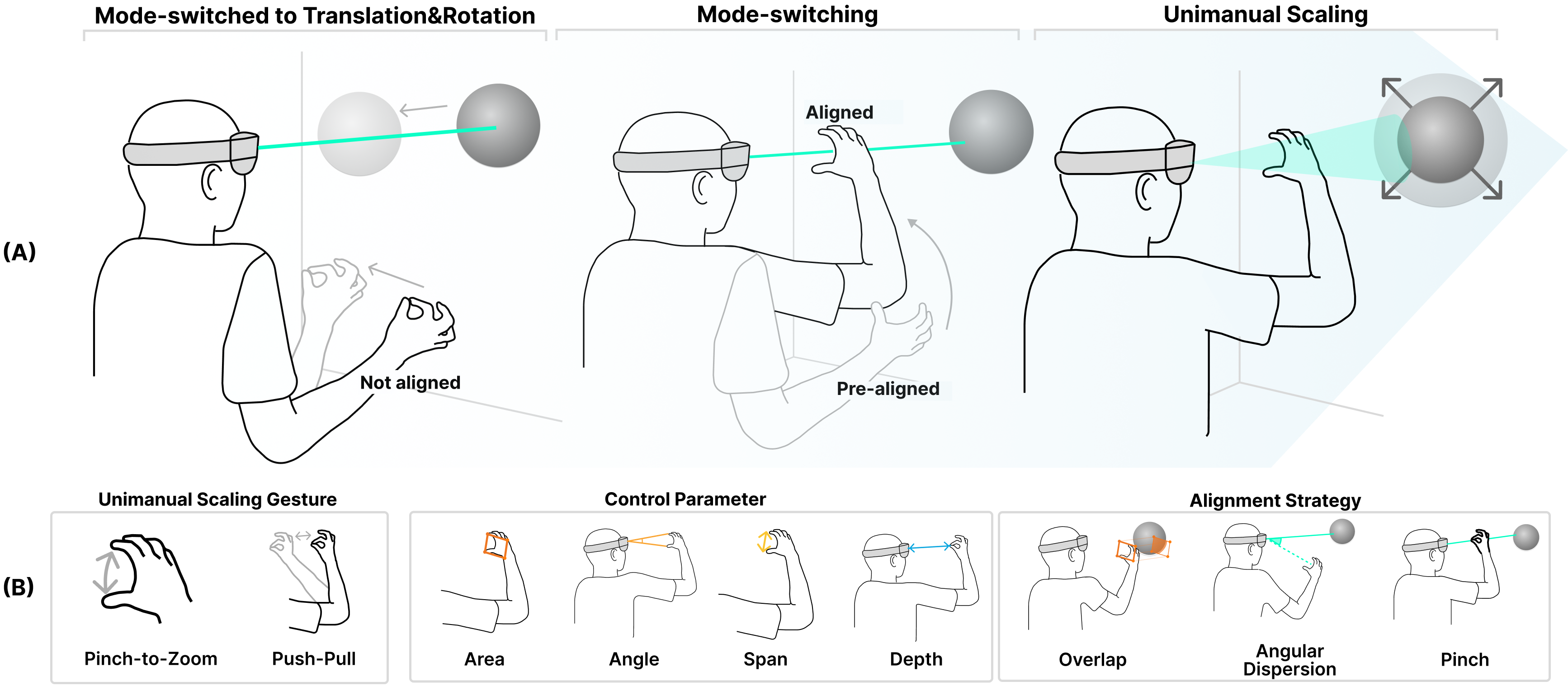}
  \caption{We propose a unimanual object manipulation by enabling single-handed scaling with gaze-hand-object alignment-based mode-switching. When the hand, gaze, and virtual object do not align, the Gaze+Pinch-based translation mode is activated (A-left). On the other hand, when the hand, gaze, and the object are aligned, in line-of-sight (A-middle), scaling mode is activated (A-right) with unimanual scaling gestures (B-left), by utilizing control parameters to calculate scaling factors (B-middle). Utilizing the proposed gaze-hand-virtual object alignment strategies (B-right) enables us to perform all manipulation features using a single hand.
  }
  \Description{}
  \label{fig:teaser}
\end{teaserfigure}
  
\maketitle

\section{Introduction}
With the advancement and emergence of extended reality (XR) and glasses-based form factors like the Meta Ray-Ban Display, Snap Spectacles, and XReal Pro, researchers and practitioners are envisioning their use in everyday contexts~\cite{grubert16, Manakhov24, Rasch25}. To achieve this goal, a key interaction capability is to operate the interface with only a single hand~\cite{karlson2008understanding}.
According to Microsoft's Persona Spectrum, unimanual interactions are essential for permanent or temporary impairments of the upper limbs, as well as for situational impairments such as a parent holding a baby \cite{microsoft2016inclusive}. One-handed situations are also common in daily life such as when multitasking with a pen, navigating on-the-go with a bag, or simply holding everyday objects like a cup \cite{ng2013impact, karlson2008understanding}. 
Accordingly, for mobile phones, alternative unimanual or single-finger techniques have been explored when established bimanual or multi-finger approaches exist \cite{holman2013unifone, esteves2022one, boring2012fat}.
Given the nature of wearable devices \cite{pascoe2000using}, the same necessity of alternative unimanual interaction also applies to XR.

However, unimanual control remains limited in current XR hand interactions. 
To operate XR user interfaces (UIs), devices increasingly support eye-gaze as a fast pointing mechanism in the scene, coupled with pinch gestures for object manipulation. Users select and move objects by looking at them and performing a pinch with the index finger and thumb. Similarly, object rotation is supported by default with one hand through rotating the hand. In contrast, to scale an object, users pinch with both hands to change the distance between them. This is a standard model for atomic rotate-scale-translate (RST) tasks as proposed in prior scientific research \cite{pfeuffer2017gaze,chatterjee15} and used, e.g., in the Apple Vision Pro XR headset. Yet, for basic scaling tasks such as resizing a window, adjusting the size of an image, or zooming within a map, there is currently no default option for one-handed control. 

Therefore, we explore a mode-switching method, Align-to-Scale, that exploits a particular eye-hand spatial coordination pattern to enable unimanual manipulation in XR. The idea is to activate the scaling mode when the user's hand and gaze are aligned (Fig.~\ref{fig:teaser}(A-right)), and deactivate when they are separate~(Fig.~\ref{fig:teaser}(A-left)). This is inspired by the gaze-hand alignment concept used for a selection~\cite{lystbaek2022gaze}, which we adopt to a framework where the aligned state serves as a cue for mode-switching from translate-rotate to scaling operations. The main advantage is that, instead of switching between different hands, users can perform the whole manipulation with the same hand, enabling complete one-handed object manipulation. 

The main research question is how users perform unimanual scaling and mode-switching with Align-to-Scale. 
We designed four unimanual scaling techniques by combining gesture type, gaze-hand alignment strategies, and control parameters. For scaling gestures, we adopted familiar Pinch-to-Zoom (varying the distance between thumb and index) and Push-Pull (moving the hand back and forth) (Fig. \ref{fig:teaser}(B-left)). Secondly, we examine alignment strategies for users to robustly perform mode-switching~(e.g., Overlap, Angular Dispersion, and Pinch) (Fig. \ref{fig:teaser}(B-right)). The conditions are studied in a translate-scale task, where users first move an object, then switch mode, and scale the target. 

Our results indicate that, despite performance costs associated with transitioning from bimanual to unimanual, participants were able to understand and execute unimanual manipulation and alignment-based mode-switching. Each technique had pros and cons regarding robust mode separation, scaling accuracy, and perceived intuitiveness. Here, we emphasize that our comparison with the baseline of bimanual scaling is not about outperforming, but providing qualitative benefits of an alternative unimanual option. Based on the results, we derive design guidelines for selecting the most suitable unimanual scaling gesture, alignment strategy, and control parameter based on user contexts and tasks, and suggest potential directions for improvement. Furthermore, we open-sourced our implementations on GitHub. Our contributions are as follows:

\begin{itemize}   
    \item We present interaction techniques that integrate gaze and hand as a mode-switching cue. Extending the familiar scaling gestures, we provide alternative unimanual scaling methods to support one-handed manipulation in XR.
    \item We present a user study evaluating the techniques in a translation-scaling task, revealing trade-offs between mode-switching and scaling performance, and perceived intuitiveness, alongside the performance cost of transitioning from bimanual to unimanual.     
    \item We present guidelines for the proposed method, informing the techniques' suitability for practical use cases.
\end{itemize}

\section{Related Work}
\subsection{Mode-switching Techniques for XR hand interactions} \label{sec:rw_modeswitch in xr}
Bimanual mode-switching is known to be efficient, due to its asymmetric division of labor where the non-dominant hand provides mode-switching cues while the dominant hand does the primary task \cite{guiard1987asymmetric}. Examples are raising the non-dominant hand~\cite{kim2014intuitive, tan2013informatics, vatavu2013comparative}, making a gesture \cite{hayatpur2019plane, kim2018agile}, or touching a body part with it \cite{walter2013strikeapose, locken2012user,hayashi2014wave}.
For unimanual mode-switching, performing distinctive gestures \cite{freeman2012freehand, vogel2005distant, kim2025t2iray, hayatpur2019plane}, making a pinch with different fingers \cite{shi2024experimental}, and moving or rotating the hand \cite{smith2019experimental} were suggested.
Surale et al. compared hand-based mode-switching in VR and reported that bimanual mode-switching was less error-prone than unimanual mode-switching. They also identified that making a pinch with different fingers is a promising option, whereas distinctive hand gestures for mode-switching cause user confusion~\cite{surale2019experimental}. However, their experiment was limited to a single task type, line drawing, lacking practical insights on whether unimanual mode-switching can also support different manipulation modes, including scaling, which we aim to explore. 

These hand-based mode-switching also suffer from increased cognitive load as the number of gestures and complexity increase~\cite{laviola20173d}. Thus, there have been approaches to introduce new modalities alongside hand input, including feet \cite{velloso2015feet}, voice \cite{bolt1980put}, head \cite{shi2021exploring}, or using external tools \cite{stoakley1995virtual, kim2023vibaware}. In this work, we focus on gaze, a modality that has become a common input in XR. Recent work has proposed the alignment between gaze and hand in 3D space as a new interaction cue \cite{lystbaek2022gaze}. This gaze-hand alignment has been used for UI \cite{lystbaek2022gaze, wagner2023fitts} or region selection \cite{shi2023exploring}, or accessing remote objects \cite{liu2025glance}.
Building on this, as shown in Tab. \ref{tab:interaction_comparison}, we address that our alignment-based mode-switching is the first attempt to integrate gaze and hand for mode-switching to enable unimanual scaling. 

\subsection{Unimanual Object Manipulation in XR}
Unimanual interaction enables multitasking~\cite{boring2012fat, li2013bezelcursor}, minimizes attentional load~\cite{pascoe2000using, karlson2007thumbspace} and lowers barriers of acceptability \cite{serrano2014exploring} and accessibility~\cite{yamagami2022two}. 
However, current scaling in XR is done by varying the distance between two hands \cite{song2012handle, pierce1997image, lee2024bimanual} based on a metaphor of `stretch and squeeze' \cite{mendes2019survey, van19971997}.

In case of unimanual scaling, based on a natural tendency to use symmetrical thumb-index movement for resizing \cite{brouet2013understanding}, a Pinch-to-Zoom (PTZ) has been utilized in 2D UI~\cite{avery2014pinch,hinckley1998interaction, kaser2011fingerglass, malacria2010clutch}. However, when PTZ is applied within the Gaze+Pinch framework, gesture misclassifications can occur between the full pinch gesture and the PTZ. To avoid this, Dewitz et al. employed non-dominant hand pinching to activate PTZ, making the interaction bimanual \cite{dewitz2021virtuality}. 
Another option is `Push-Pull' gesture, which involves moving the hand back and forth ~\cite{buschel2019investigating, yoo20103d, stellmach2012investigating, nancel2011mid}. Stellmach et al. implemented it by raising the non-dominant hand \cite{stellmach2012investigating}, and Yoo et al. required users to make a distinctive gesture \cite{yoo20103d} when making the Push-Pull gesture.
A double-pinch gesture was also used for zooming, and a normal pinch for panning \cite{buschel2019investigating}, yet the double-pinch is only compatible with Push-Pull and not PTZ, as it requires the thumb and index finger to be in contact. 

In Tab. \ref{tab:interaction_comparison}, we compared previous mode-switching methods for supporting these unimanual gestures. Hand Position/rotation-based mode-switching was the only method compatible with both. Shi et al. used wrist rotation to switch between selection modes; however, it was physically uncomfortable \cite{shi2024experimental}. Smith et al. investigated hand-depth-based switching during translation tasks, which proved to be the fastest but also the most inaccurate \cite{smith2019experimental}. Therefore, building on these approaches, we adopt the gaze-hand alignment concept as a mode-switching to enable these intuitive unimanual gestures in XR. As mode-switching mostly refers to an implicit cue rather than an explicit visual UI \cite{raskin2000humane}, and also considering its visually disruptive manner \cite{kurtenbach1994user, guimbretiere2000flowmenu}, we do not consider visual UI-based unimanual manipulations like 3D bounding boxes \cite{microsoft_boundingbox}.

\begin{table}[h]
  \centering
  \caption{Comparison of Align-to-Scale with previous mode-switching methods. Methods are analyzed based on their support for unimanual interactions, integration of gaze and hand modalities, and compatibility with scaling gestures. A $\checkmark$ indicates that the mode-switching method supports the dimension, while a - indicates a lack of support.}  
  \label{tab:interaction_comparison}
  \small 
  \setlength{\lightrulewidth}{0.01em} 
  \setlength{\tabcolsep}{4pt}

  \begin{tabular}{p{3.2cm} c c c c c} 
    \toprule
    \multirow{2}{*}[-0.5ex]{Mode-switching} & \multirow{2}{*}[-0.5ex]{Unimanual} & Integration of & \multicolumn{3}{c}{Gesture Compatibility} \\
    \cline{4-6}
    & & Gaze \& Hand & PTZ & Push-Pull & Bimanual \\
    \midrule
    
    Non-Dominant Hand & \multirow{2}{*}{-} & \multirow{2}{*}{-} & \multirow{2}{*}{$\checkmark$} & \multirow{2}{*}{$\checkmark$} & \multirow{2}{*}{-} \\ 
    \cite{stellmach2012investigating, dewitz2021virtuality} & & & & & \\ \midrule    

    Distinctive Gesture & & & & & \\ 
    \cite{hayatpur2019plane, yoo20103d} & $\checkmark$ & - & - & $\checkmark$ & - \\ \midrule
    
    Different Finger Pinch & & & & & \\ 
    \cite{surale2019experimental, shi2024experimental} & $\checkmark$ & - & - & $\checkmark$ & - \\ \midrule
    
    Hand Position/Rotation & & & & & \\ 
    \cite{shi2024experimental, smith2019experimental} & $\checkmark$ & - & $\checkmark$ & $\checkmark$ & - \\ \midrule
    
    Bimanual Scaling & & & & & \\ 
    \cite{lee2024bimanual, song2012handle} & - & - & - & - & $\checkmark$ \\ \midrule

    Double-Pinch & \multirow{2}{*}{$\checkmark$} & \multirow{2}{*}{-} & \multirow{2}{*}{-} & \multirow{2}{*}{$\checkmark$} & \multirow{2}{*}{-} \\ 
    \cite{buschel2019investigating} & & & & & \\ 
    
    \specialrule{0.06em}{0.5ex}{0.5ex} 
    \textbf{Align-to-Scale} & \multirow{2}{*}{\boldmath$\checkmark$} & \multirow{2}{*}{\boldmath$\checkmark$} & \multirow{2}{*}{\boldmath$\checkmark$} & \multirow{2}{*}{\boldmath$\checkmark$} & \multirow{2}{*}{-} \\ 
    \textbf{Mode-switching (Ours)} & & & & & \\ 
    \specialrule{0.06em}{0.5ex}{0ex} 
  \end{tabular}
\end{table}

\section{Design Space for Unimanual Object Manipulation}
\label{sec:design space}
\begin{figure}[h]
    \includegraphics[width=\textwidth]{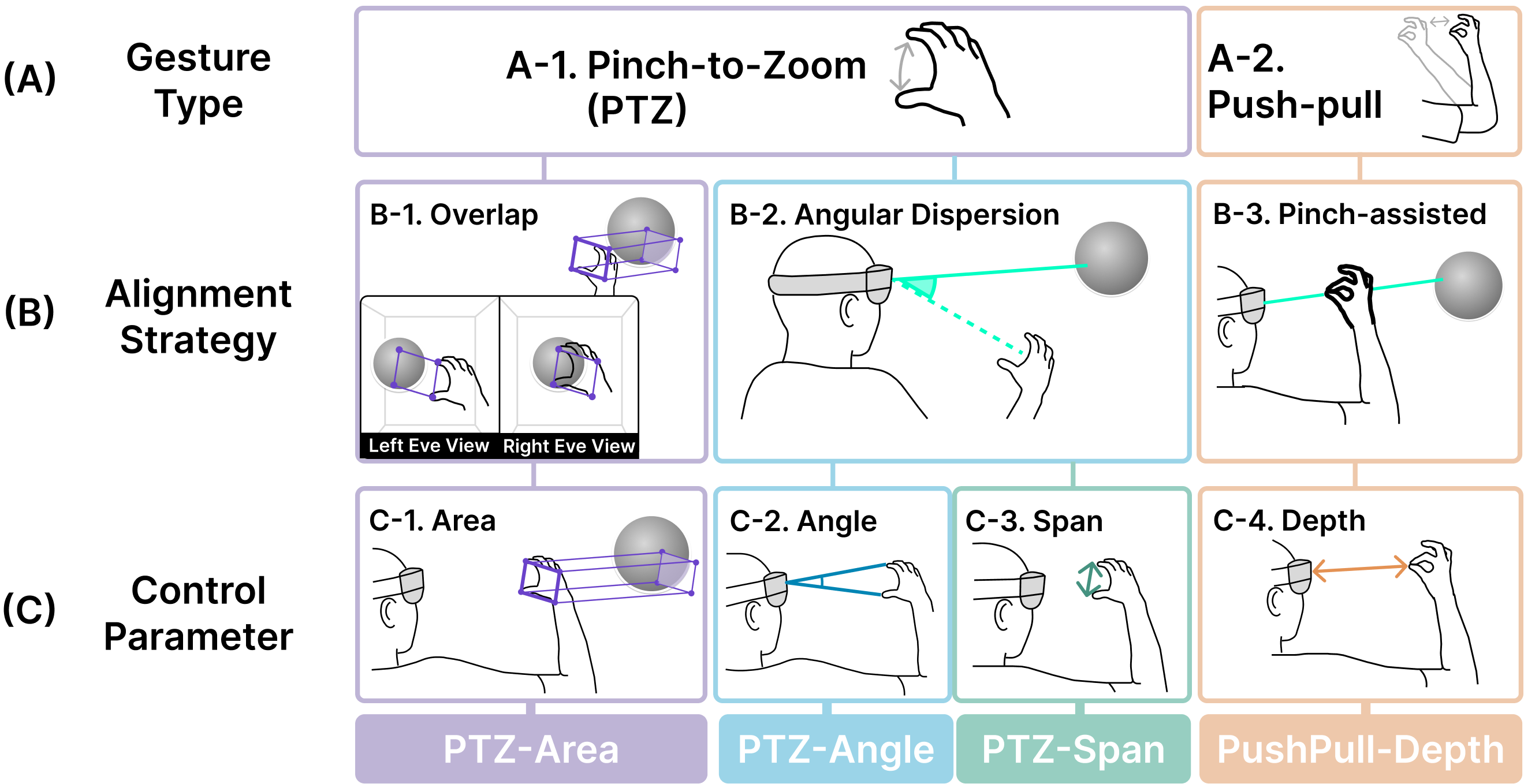}
    \caption{Design space of unimanual scaling interactions. We derived four interactions: PTZ-Area, -Angle, -Span, and Push-Pull-Depth, by combining design factors of scaling gesture type, alignment strategy, and control parameter. We chose two unimanual scaling gestures: Pinch-to-Zoom (PTZ) (A-1) and Push-Pull (A-2). Three gaze-hand alignment strategies for mode-switch: stereoscopic view area overlap (purple rectangular area, B-1), angular dispersion between gaze–object and gaze–hand (cyan angle, B-2), and pinch gestures performed when the gaze, hand, and object were aligned(B-3) were tested. (C) Each input depicted in purple, blue, green, and orange represents control parameters used to compute scaling factors.}
    \label{fig:space}
\end{figure}

The design space for unimanual object manipulation involves three design factors: \textit{gesture type}, \textit{alignment strategy}, and \textit{control parameter} (Fig. \ref{fig:space}). 
The \textit{alignment strategy} is a factor that decides how to integrate gaze and hand. Through it, we aim to fill in the gap of previous mode-switching research, alongside \textit{gesture type} and \textit {control parameter} selected as representative factors highly relevant to scaling interactions \cite{schubert2023intuitive, nancel2011mid}.   
By combining elements from each factor, we derived four interactions: PTZ-Area, PTZ-Angle, PTZ-Span, and Push-Pull-Depth. All interaction parameters, including thresholds, are chosen via pilot testing and are detailed in Tab. \ref{tab:variables}. All subsequent mentions of the variables refer to those in this table. 
For reproducibility, we open-sourced the implementations \footnote{\url{https://github.com/MinKim242/ETRA_Align-to-Scale.git}}. 

\subsection{Gesture Type}
\textbf{Pinch-to-Zoom (PTZ).} PTZ gesture is a familiar unimanual zooming gesture on 2D UI. Increasing the span between the thumb and index finger tips scales the object up, while decreasing it scales the object down (Fig. \ref{fig:space}(A-1)). However, it has not been adopted in XR due to an absence of a robust mode-switching method that distinguishes PTZ from the normal pinch gesture for selection or translation in Gaze+Pinch. Thus, we propose utilizing gaze-hand alignment as a mode-switching cue; when aligned, thumb and index movement is mapped to scaling through PTZ, and when not aligned, their movement would be interpreted as the normal pinch. 

\textbf{Push-Pull.} Moving the hand back-and-forth corresponds to pushing the object away to make the size smaller and pulling it closer to the user's body to make it bigger (Fig. \ref{fig:space}(A-2)). Unlike PTZ, this gesture is compatible with unimanual hand-based mode-switching (Distinctive Gesture, Different Finger Pinch, or Double-Pinch in Tab. \ref{tab:interaction_comparison}). Thus, we compare the two gestures to explore the trade-off between the intuitiveness of the gesture and the robustness of mode-switching.

\subsection{Alignment Strategy}
Alignment Strategy denotes the type of cue to determine whether the hand is in line-of-sight with gaze and the object. 
We derived three variations of alignment strategy methods to distinguish between normal and scaling modes.

\textbf{Overlap.} We used an overlap-based gaze-hand alignment (Fig. \ref{fig:space}(B-1)) with the PTZ gesture. Mode-in to scaling is triggered when the stereoscopic view area overlaps with the object, and while being gazed at. The stereoscopic view area is defined as the rectangular region formed by projecting the thumb and index fingertips from the left and right eyes. We calculate how much of the view area the object covers, and vice versa. If either ratio exceeds its overlap threshold, the system assumes the object and the view area are overlapping. The stereoscopic view area is illustrated as a purple rectangular area, while the overlap is denoted by a purple fill inside the view area (Fig. \ref{fig:space}(B-1)). 

The overlap-based alignment is driven by two motivations. First, we expect the method to reproduce the sensation of direct scaling in XR, as when performing PTZ on 2D screens \cite{hinckley1998interaction, pfeuffer2016partially}. Second, we aim to improve the conventional Head Crusher technique, which utilizes a similar concept of framing the target within the space between fingers \cite{pierce1997image}. Prior experiments indicated the conventional logic of ray-casting from eye through the thumb-index midpoint resulted in a low selection performance \cite{wagner2023fitts}. 

\textbf{Angular Dispersion.} It is a conventional cue of gaze-hand alignment \cite{lystbaek2022gaze}. The angular dispersion is measured by the angle between a gaze ray and a vector from the gaze origin to the hand. The scaling mode is activated when the angular dispersion between gaze and the hand falls below a mode-in threshold (cyan angle in Fig. \ref{fig:space}(B-2)), and the mode-out threshold was set bigger than the mode-in threshold to prevent unintentional mode-outs. 

\textbf{Pinch-assisted.} We adopted pinch-assisted gaze-hand alignment (Fig. \ref{fig:space}(B-3)) as a representative from hand gesture-based mode-switching \cite{surale2019experimental}. When users make a pinch with their hands raised near the gaze, mode-in to scaling occurs. Moving the hand for-/backward will then scale. When they make a pinch gesture without the alignment, translation mode is triggered. The same thresholds for mode-in and out were used with the angular dispersion. 

\subsection{Control Parameter}
The scaling factor is a multiplier that determines how much an object's size changes relative to its original size. The scaling factor is calculated as a ratio between the new size and the initial object size. 
We calculated $s_t$, the current scale of the object at time $t$, with the following equation of $s_t = s_0 \times \frac{I_t}{I_0}$. $s_0$ is an object scale at the start of the scaling mode, $I_t$ is a current input from the scaling gesture at time $t$, and $I_0$ is an initial input at the start of the scaling mode. Here, the scaling factor is $\frac{I_t}{I_0}$. By the term control parameter, we refer to the input $I$ used for the computation of the scaling factor. 

By proposing \textbf{stereoscopic view area} (Fig. \ref{fig:space}(C-1)) and \textbf{angle} as control parameters, we explore a new style of scaling where scaling is influenced by both a span between the thumb and the index tips, and the depth of the hand. The same stereoscopic view area used for overlap-based alignment is reused as a control parameter. The angle for the scaling factor refers to the angle between two vectors from the gaze origin to the tips of the thumb and index finger, respectively. Both the overlap area and the angle increase as the finger span widens or the hand moves closer in depth, resulting in a scale-up, and a scale-down for the opposite direction. We also included \textbf{Span}-only as a familiar control parameter used in 2D PTZ  (Fig. \ref{fig:space}(C-3)). For the Push-Pull scaling, we implement the \textbf{depth} of the hand as a control parameter (Fig. \ref{fig:space}(C-4)). The depth of the hand is defined as the perpendicular distance from the head to the hand. 

To prevent noises in input and ensure a comfortable range of hand movements \cite{fuente2010user}, the control parameters were clamped between minimum and maximum values of each input. 

\begin{table}[h]
  \centering
  \caption{Interaction parameters used in the study. All the mentioned thresholds in Sec. \ref{sec:design space} are listed here. Note that the unit of the minimum and maximum clamping for PTZ-Area is a proportion of the stereoscopic view area on the HMD display between 0 to 100\%.}
  \label{tab:variables}
  \footnotesize 
  \setlength{\tabcolsep}{5pt} 
  
  \begin{tabular}{p{2.5cm}|p{2cm}|p{2.9cm}|p{3.8cm}}
    \hline
    \textbf{Design Factor} & \textbf{Technique} & \textbf{Parameter} & \textbf{Value} \\ 
    \hline
    Alignment Strategy: Overlap & PTZ-Area & Overlap Threshold & Min. 25\% of Stereoscopic View Area OR Min. 15\% of Object is covered by each other \\
    \hline
    
    \multirow{2}{*}{\begin{tabular}[c]{@{}l@{}}Alignment Strategy: \\ Angular dispersion, \\ Pinch-assisted\end{tabular}} 
    & \multirow{2}{*}{\begin{tabular}[c]{@{}l@{}}PTZ-Angle, \\ -Span, \\ Push-Pull-Depth\end{tabular}} 
    & \multirow{1}{*}[-0.7ex]{Mode-in Threshold} & \multirow{1}{*}[-0.7ex]{15°} \\[1.5ex] \cline{3-4} 
    
    & & \multirow{1}{*}[-0.7ex]{Mode-out Threshold} & \multirow{1}{*}[-0.7ex]{17°} \\[1.5ex]    

    \hline
    \multirow{5}{*}{Control Parameter} 
    & PTZ-Area & \multirow{5}{*}{Min. \& Max. of Clamping} & 0.1\%, 100\% \\ \cline{2-2} \cline{4-4}
    & PTZ-Angle & & 3°, 40° \\ \cline{2-2} \cline{4-4}
    & PTZ-Span & & 0.01m, 0.15m \\ \cline{2-2} \cline{4-4}
    & Push-Pull-Depth & & 0.1m, 0.5m \\ \cline{2-2} \cline{4-4}
    & Bimanual & & 0.01m, 0.8m \\ 
    \hline
  \end{tabular}
\end{table}

\section{User Study}
\subsection{Task Design}
We conducted a within-subjects design to assess the performance and experience of the interactions. Each trial required participants to first translate a virtual object to a specified target position, mode-switch to scaling mode, and then scale the object to match a target size. Our analysis focused on mode-switching events: mode-in to translation, mode-in to scaling, and mode-out from scaling mode to idle state. The mode-out for translation was excluded as it occurs with the instant release of the pinch. The rotation task was excluded because it occurs within the same mode as translation for hand interactions  \cite{mendes2019survey}, requiring no explicit mode-switch. Standard frameworks like Virtual Hand \cite{laviola20173d}, and Gaze+Pinch \cite{pfeuffer2017gaze} also combine both tasks into a single mode.

There were three independent variables used in the study: 5 techniques (including the baseline), 4 target positions (up, down, left, right), and 4 target scales ($\times 0.4$, $\times 0.67$, $\times 1.5$, $\times 2.5$). We position bimanual scaling as a relevant baseline for both scaling and mode-switching, as use of the non-dominant hand was also frequently suggested as a cue for mode-switching \cite{surale2019experimental, ruiz2008model}. A single session consisted of 16 combinations of target scales and positions, and participants repeated three sessions, with the first session as a practice. Thus, only the last two sessions were used, and the total number of trials for analysis was 160 (5 techniques $\times$ 4 target scales $\times$ 4 target positions $\times$ 2 sessions). The order of the techniques was randomized among participants to prevent learning effects. 

\subsection{Implementation}
The study environment was developed in Unity 2022.3.39f1 with Varjo XR-3 HMD (90Hz, 115$^{\circ}$ FoV) embedded with an eye tracker (200 Hz). We used Ultraleap SDK's pinch detection, and the hand-tracking data was smoothed with a 1\texteuro~filter \cite{casiez20121}, with parameters empirically set at $f_{c_{min}} = 1.0$ and $\beta = 90$ \cite{wagner2024eye}. 
Following the previous works \cite{kim2025pinchcatcher}, \cite{ pfeuffer2017gaze}, we also adopted the change in the object outline color as an indicator of current mode: orange outline in translation mode, yellow in scaling mode, and white when gazed at. 

\begin{figure}[h]
    \includegraphics[width=\textwidth]{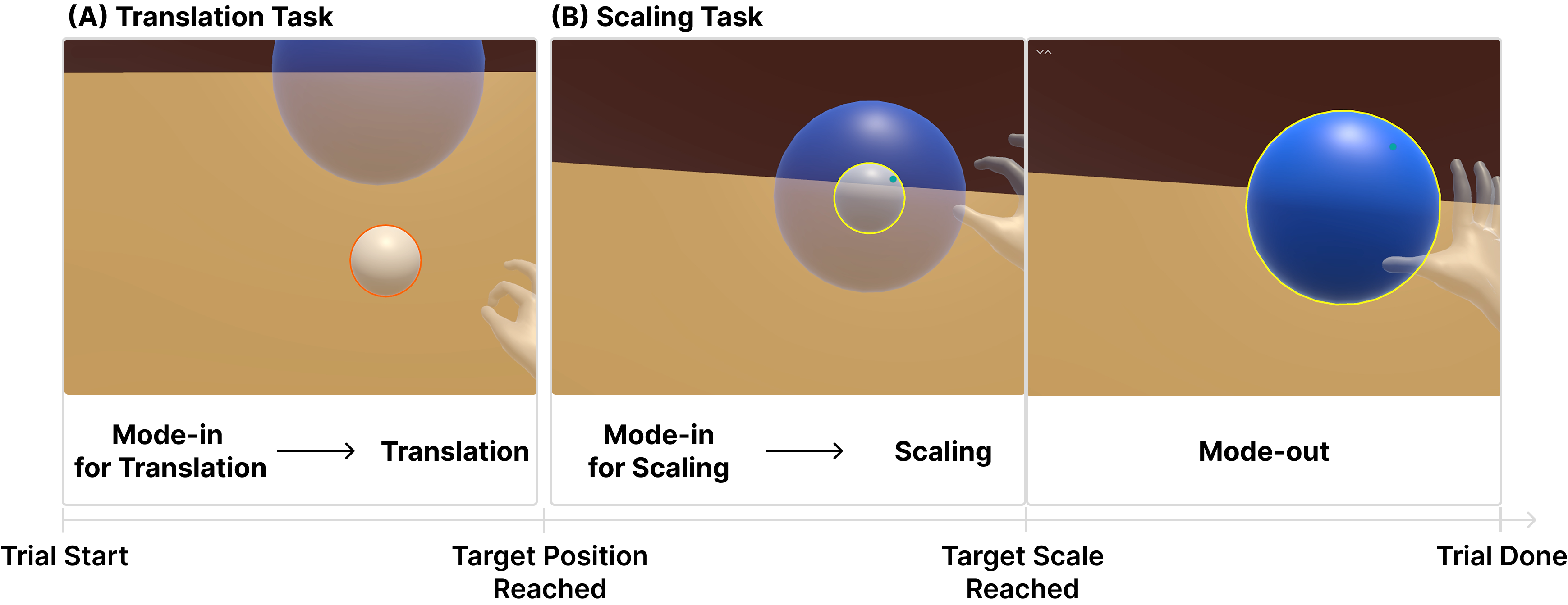}
    \caption{User study overview and temporal order of a single trial. Participants manipulated a white sphere to match the position and scale of a semi-transparent blue target sphere. Participants first translated the white sphere toward the blue sphere. The outline of the white sphere turned orange when switched to translation mode (A). Then, participants mode-in to scaling mode. When they switched to scaling mode, the outline color of the white sphere turned yellow (B-left). When the white sphere’s scale reached the target scale, it turned opaque blue, indicating that participants to switch out of the scaling mode (B-right). The trial ended when the participant mode-out of scaling.}
    \label{user study overview}
\end{figure}

\subsection{Procedure}
We recruited 20 participants $(M=24.85, SD=3.08, 7 Female)$, and none reported visual or motor impairments (i.e., color deficiencies) that would affect performing the tasks. The study was approved by the Institutional Review Board~(IRB), and participants received \$15 for their participation. After signing a consent form, we introduced interaction techniques and the tasks for 15 minutes. For each technique, participants completed one practice session followed by two main sessions, all of them with the same number of trials. Each main session was followed by subjective ratings and 5 minutes rest. After all the sessions, participants ranked the five techniques from top to bottom and took a short interview.
 
During the main sessions, participants were shown two semi-transparent sphere-shaped objects (Fig. \ref{user study overview}). The white sphere was the object to manipulate, while the blue sphere indicated a target position and scale. All the spheres appeared from 2~m in depth from the participant. At the beginning of each trial, the white sphere that participants had to manipulate appeared in front of them. The blue sphere was placed at an angular offset of $35^\circ$ from the participant’s front direction, located in one of four directions~(target positions). The white sphere had a diameter of $14^\circ$, and the blue sphere was scaled to one of the four target scales.

Participants first started with the translation task (Fig.~\ref{user study overview}(A)). They translated the white sphere to the position of the blue sphere using Gaze+Pinch~\cite{pfeuffer2017gaze}. The sphere snapped to the position when its center was within 0.15~m of the target position to indicate task completion and start scaling the task. The snapping feature was utilized to ensure center-to-center alignment between the white sphere, which participants are controlling, and the blue sphere indicating the target scale. 
Then, participants started scaling the white sphere using different techniques assigned for each session (Fig.~\ref{user study overview}(B-left)). The color of the white sphere changed to blue at the moment the sphere reached the target scale (Fig.~\ref{user study overview}(B-right)). This informs participants to mode-out of the scaling mode and move on to the next trial. We considered the target scale to be achieved once the diameter difference between the white and blue spheres was under 0.1. In the case of the scaling task, participants had to stop the scaling on their own while trying to keep the scale of the sphere as close to the target scale. To measure mode-switching time, we controlled the scaling task such that participants could only perform a single scaling attempt without clutching. Trials in which participants failed to reach the target scale on the first try were recorded as scaling error trials. 

\subsection{Evaluation Metrics}
We have two error categories: mode-switching and scaling errors. Mode-switching errors are divided into mode-in errors for translation and scaling tasks. A mode-in error is a transition to a mode that does not align with the current task, including gesture misclassifications between the pinch gesture for translation and PTZ. Specifically, during translation, errors are logged when the system switches to scaling instead of translation, and vice versa for scaling tasks. Scaling errors occur when participants fail to reach the target scale. Trials are marked as 1 if an error occurred and 0 otherwise. The error rate is computed by averaging across trials for each technique \cite{surale2017experimental, surale2019experimental}. We also calculated the overall mode-switching error rate by logging the error trials regardless of the mode-in error types. We converted the error rate to a percentile, and a closer to 100 \% indicates a greater likelihood of committing that error type.

We also measure mode-in time, defined as the duration to transition from the idle state to the mode corresponding to the current task. For scaling performance, mode-out time, which is the duration from scaling mode back to idle, was analyzed, as maintaining a stable object scale during this period is critical to achieving the intended scaling. Furthermore, we measure the scaling difference, defined as the deviation between the final object scale after mode-out and the target scale. Lastly, subjective evaluations were conducted using NASA Task Load Index~(NASA-TLX)~\cite{hart2006nasa}, our own usability questionnaires, and preference rankings on the techniques. We used a raw score of NASA-TLX on a 7-point scale following previous experiments on XR gaze interactions \cite{wagner2023fitts, yu2021gaze}. 

The overall mode-switching error rate from mode-in errors to translation and scaling, mode-in error rates and time for each task were used to assess the robustness of mode-switching. Scaling error rate, scale difference, and mode-out time for scaling were used to evaluate the scaling performance. 

\section{Result}
While all trials were included in the error rate analysis, analyses of completion time and scale difference excluded trials that failed to reach the target scale. We removed 50 trials (1.8\%) identified as outliers due to tracking loss, exceeding 3 SDs from the mean task completion time calculated for each condition. The Shapiro-Wilk test indicated non-normal distributions. Thus, we conducted a nonparametric two-way repeated-measures ANOVA using the Aligned Rank Transform (ART) \cite{wobbrock2011aligned}, with Technique and Target Scale as within-subject factors. When significant main or interaction effects were found, post-hoc pairwise comparisons were performed using the ART-C procedure\cite{elkin2021aligned} with Holm correction. For brevity, we only report the significant pairs within the same Technique when interaction effects are significant. However, we included both Technique and the Technique $\times$ Target Scale interaction in our figures to provide an overview. Detailed results are included in the supplementary material.

\begin{figure}[h]
    \includegraphics[width=\textwidth]{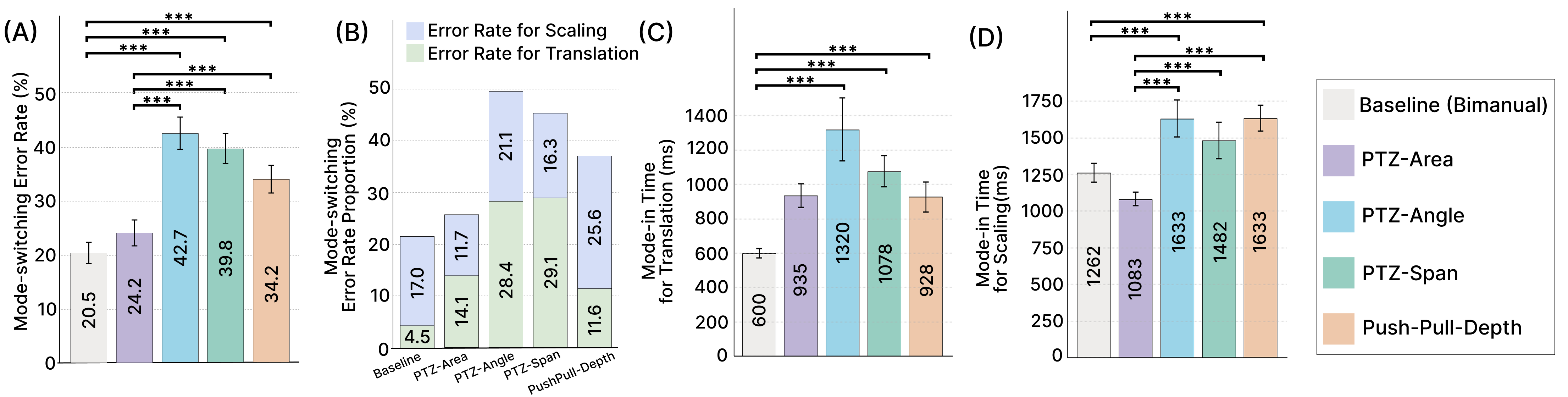}
    \caption{Results on mode-switching performance, indicating robust distinction between modes. (A) Overall mode-switching error rates by Technique. Overall mode-switching error includes both the mode-in for translation and the scaling mode. Error rates closer to 100\% indicate more frequent incorrect mode activations. (B) Proportions of mode-switching error rates by Technique. Multiple error types can occur from a single trial. (C) The mean mode-switching time it took for mode-in to translation. (D) The mean mode-switching time during the mode-in for scaling mode. Significance levels are indicated as *p < .05, **p < .01, ***p < .001; error bars represent standard error.    
}
    \label{fig:result1}
\end{figure}

\subsection{Mode-switching Performance}
\subsubsection{Overall Mode-Switching Error Rate} The Overall Error Rate was significantly influenced only by the Technique factor ($F(4, 361)=23.0,\ p<.001,\ \eta_p^2=.20$) (Fig.~\ref{fig:result1}(A)). The post-hoc test revealed that Bimanual scaling ($M=20.5,\ SD=17.5$) resulted in a lower Overall Mode-switching Error Rate than PTZ-Angle ($M=42.7,\ SD=26.5;\ t(361)=-7.58,\ p<.001,\ d=0.80$), PTZ-Span ($M=39.8,\ SD = 25.1;\ t(361)=-7.03,\ p<.001, \ d=0.74$), and Push-Pull scaling ($M=34.2,\ SD = 22.8;\ t(361)=-5.26,\ p<.001,\ d=0.55$). PTZ-Area ($M=24.2,\ SD=21.5$) also led to a lower Overall Mode-switching Error Rate than PTZ-Angle ($t(361)=-6.18,\ p<.001,\ d=0.65$), PTZ-Span ($t(361)=-5.63,\ p<.001,\ d=0.59$), and Push-Pull scaling ($t(361)=-3.86,\ p<.001,\ d=0.41$). 

\subsubsection{Mode-in Error Rate for Translation}
The Technique was the only significant factor ($F(4,361)=38.5,\ p<.001,\ \eta_p^2=.30$) (Fig.~\ref{fig:result1}(B)). Post-hoc comparisons revealed that PTZ-Angle ($M=28.4$, $SD=27.2$) and PTZ-Span ($M=29.1$, $SD=25.6$) yielded significantly higher error rates than all other techniques. PTZ-Angle had significantly higher error rates than the baseline ($M=4.5,\ SD = 8.0;\ t(361)=9.60,\ p<.001, \ d=1.01$), Push-Pull-Depth ($M=1.6,\ SD = 16.7;\ t(361)=-6.04,\ p<.001, \ d=0.64$), and PTZ-Area ($M=14.1,\ SD = 18.2;\ t(361)=5.52,\ p<.001, \ d=0.58$). PTZ-Span led to significantly higher error rates than the baseline ($t(361)=9.60,\ p<.001, \ d=1.01$), Push-Pull-Depth ($t(361)=6.86,\ p<.001, \ d=0.72$), and PTZ-Area ($t(361)=6.33,\ p<.001, \ d=0.67$). Additionally, the baseline resulted in significantly lower error rates than PTZ-Area ($t(361)=-4.08$, $p=.001,\ d=0.43 $) and Push-Pull-Depth ($t(361)=-3.56$, $p<.001,\ d=0.37$).

\subsubsection{Mode-in Error Rate for Scaling}
There was significant main effect of Technique ($F(4,361)=7.41,\ p<.001,\ \eta_p^2=.076$) and Target Scales ($F(3, 361)=4.34,\ p=.005$). The interaction effect was also significant ($F(12, 361)=3.40,\ p<.001,\ \eta_p^2=.10$) (Fig. \ref{fig:result1}(B)).
The significant pairs of interaction effects within the same Technique were observed for PTZ-Span. Specifically, the Target Scale of $\times 2.5$ yielded significantly higher values than $\times 0.4$ ($t(361) = -4.96$, $p < .001$, $d=0.52$) and $\times 0.67$ ($t(361) = -4.21$, $p = .006$, $d=0.44$).

\subsubsection{Mode-in Time for Translation}
We found a significant effect of Technique on the metric (Fig.~\ref{fig:result1}(C)) ($F(4,361)=17.1,\ p<.001,\ \eta_p^2=.16$). From the post-hoc test, the Bimanual method ($M=600$, $SD=251$) resulted in significantly faster mode-in time compared to other techniques; Push-Pull-Depth ($M=928$, $SD=773$, $t(361) = -5.27$, $p < .001$, $d=0.55$), PTZ-Angle ($M=1320$, $SD=1636$, $t(361) = -7.31$, $p < .001$, $d=0.77$), PTZ-Span ($M = 1078$, $SD = 800$, $t(361) = -6.92$, $p < .001$, $d=0.73$), and PTZ-Area ($M = 935$, $SD = 624$, $t(361) = -5.34$, $p < .001$, $d=0.56$).

\subsubsection{Mode-in Time for Scaling}
Main effects were significant, on Technique ($F(4,361)=13.0,\ p<.001,\ \eta_p^2=.13$) and Target Scales ($F(3,361)=5.22,\ p=.002,\ \eta_p^2=.042$). The interaction effect was also significant ($F(12,361)=2.33,\ p=.007,\ \eta_p^2=.072$) (Fig.~\ref{fig:result2}(D)).
However, subsequent post-hoc tests revealed no significant pairs from the same Target Scale.

\subsection{Scaling Performance}
\subsubsection{Scaling Error Rate}
\begin{figure}[h]
    \includegraphics[width=\textwidth]{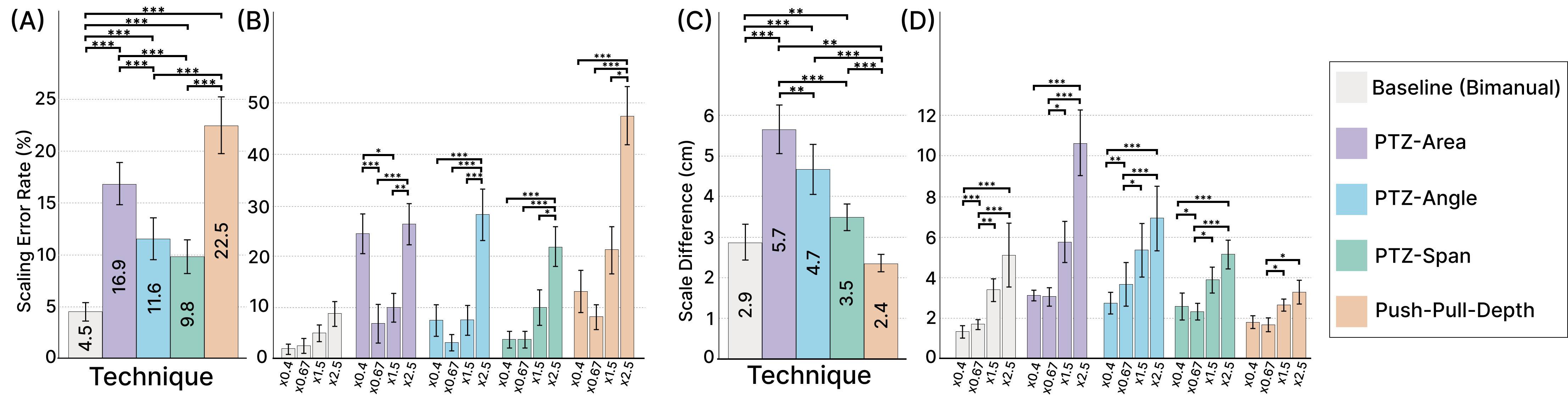}
    \caption{Results on scaling performance, showing accuracy in reaching the target scale. (A) Scaling error rate by technique. (B) Scaling error rate by Technique $\times$ Target Scale. (C) The mean scale difference shows the deviation between the final object and the target scale after mode-out. (D) The mean scale difference by Technique$\times$Target Scale. Significance levels are indicated as *p < .05, **p < .01, ***p < .001; error bars show standard error.}
    \label{fig:result2}
\end{figure}

There was a significant effect of Technique ($F(4,361)=32.8,\ p<.001,\ \eta_p^2=.27$) and Target Scale ($F(3,361)=60.4,\ p<.001,\ \eta_p^2=.33$), and interaction effect ($F(12,361)=6.72,\ p<.001,\ \eta_p^2=.18$) (Fig.~\ref{fig:result2}(A, B)). 

Except for the baseline, all the Techniques included pairs of the Target Scale with significant differences (Fig.~\ref{fig:result2}(B)). Within the PTZ-Area, $\times0.4$ led to a higher Scaling Error Rate than $\times0.67$ ($t(361)=5.41,\ p<.001,\ d=0.57$) and $\times1.5$ ($t(361)=3.86,\ p=.016,\ d=0.41$). $\times2.5$ resulted in a higher error rate than $\times0.67$ ($t(361)=5.65,\ p<.001,\ d=0.60$) and $\times1.5$ ($t(361)=4.10,\ p=.006,\ d=0.43$). For the PTZ-Angle, $\times2.5$ induced higher error rate than $\times0.4$ ($t(361)=5.19,\ p<.001,\ d=0.55$), $\times0.67$ ($t(361)=6.41,\ p<.001,\ d=0.67$), and $\times1.5$ ($t(361)=5.12,\ p<.001,\ d=0.54$). The PTZ-Span included the same significant pairs: $\times2.5$ with a higher error rate than $\times0.4$ ($t(361)=-5.31,\ p<.001,\ d=0.56$), $\times0.67$ ($t(361)=5.31,\ p<.001,\ d=0.56$), and $\times1.5$ ($t(361)=3.56,\ p=.049,\ d=0.38$). Lastly, $\times2.5$ of Push-Pull-Depth scaling also resulted in a higher error rate than $\times0.4$ ($t(361)=5.84,\ p<.001,\ d=0.61$), $\times0.67$ ($t(361)=6.90,\ p<.001,\ d=0.73$), and $\times1.5$ ($t(361)=3.76,\ p=.024,\ d=0.40$).

\subsubsection{Scale Difference}
There were significant effects of Technique ($F(4,361)=35.5,\ p<.001,\ \eta_p^2=.28$) and Target Scale ($F(3,361)=75.1,\ p<.001,\ \eta_p^2=.38$), and interaction effect ($F(12,361)=4.30,\ p<.001,\ \eta_p^2=.13$) (Fig.~\ref{fig:result2}(C, D)). 

All Techniques showed significant differences between Target Scales (Fig.~\ref{fig:result2}(D)). Within the PTZ-Area, $\times 2.5$ led to a bigger scale difference than that of $\times 0.4$ ($t(361) = 5.35$, $p < .001,\ d=0.56$) and $\times 0.67$ ($t(361) = 6.62$, $p < .001,\ d=0.70$). $\times 1.5$ showed bigger difference than the target scale of $\times 0.67$ ($t(361) = 3.68$, $p = .030,\ d=0.39$). For the PTZ-Angle, $\times 2.5$ showed a bigger scale difference than $\times 0.4$ ($t(361) = 6.38$, $p < .001$, $d=0.67$) and $\times 0.67$ ($t(361) = 5.72$, $p < .001$, $d=0.60$). Scale Difference of $\times 1.5$ was significantly bigger than $\times 0.4$ ($t(361) = 4.43$, $p = .002$, $d=0.47$) and $\times 0.67$ ($t(361) = 3.77$, $p = .022$, $d=0.40$). $\times 2.5$ of PTZ-Span resulted in bigger Scale difference than $\times 0.4$ ($t(361) = 5.46$, $p < .001$, $d=0.57$) and $\times 0.67$ ($t(361) = 5.43$, $p < .001$, $d=0.57$). $\times 1.5$  of PTZ-Span also resulted in a bigger difference than $\times 0.4$ ($t(361) = 3.71$, $p = .027$, $d=0.39$) and $\times 0.67$ ($t(361) = 3.68$, $p = .030$, $d=0.39$). The Push-Pull scaling exhibited bigger scale difference at $\times 2.5$ than $\times 0.67$ ($t(361) = 3.95$, $p = .011$, $d=0.42$). $\times 1.5$ had higher difference than $\times 0.67$ ($t(361) = 3.67$, $p = .031$, $d=0.39$). The baseline resulted in bigger scale difference at $\times 2.5$ than $\times 0.4$ ($t(361) = 6.48$, $p < .001$, $d=0.68$) and $\times 0.67$ ($t(361) = 4.85$, $p < .001$, $d=0.51$), and $\times 1.5$ higher than $\times 0.4$ ($t(361) = 5.75$, $p < .001$, $d=0.60$) and $\times 0.67$ ($t(361) = 4.12$, $p = .006$, $d=0.51$).

\begin{figure}[h]
    \includegraphics[width=\textwidth]{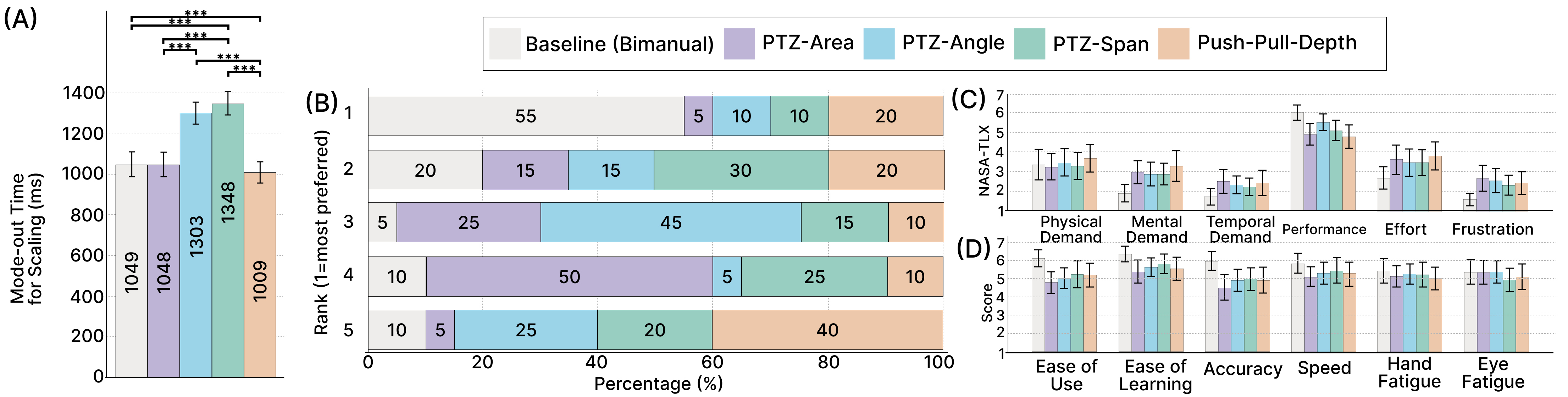}
    \caption{(A) The mean of mode-out time for scaling, the duration from scaling mode back to the idle state, by Technique. (B) Results on the ranking of user preferences. (C) Results of raw NASA-TLX score measured in 7-point scale by Techniques. (D) Subjective ratings of Techniques. Significance levels are indicated as *p < .05, **p < .01, ***p < .001; error bars represent standard error. 
    }
    \label{fig:result3}
\end{figure}

\subsubsection{Mode-out Time for Scaling}
The influences of Technique ($F(4,361)=20.3,\ p<.001,\ \eta_p^2=.18$) and Target Scale ($F(3,361)=5.83,\ p<.001,\ \eta_p^2=0.046$) were statistically meaningful. Their interaction effect was also significant ($F(12,361)=5.50,\ p<.001,\ \eta_p^2=.15$) (Fig.~\ref{fig:result3}(A)) 
There were significant pairwise differences across target scales within PTZ-Area. Specifically, the smaller target scales showed significantly faster mode-out times: $\times0.4$ was faster than both $\times1.5$ ($t(361)=-7.23$, $p<.001$, $d=0.76$) and $\times2.5$ ($t(361)=-6.06$, $p<.001$, $d=0.64$), while $\times0.67$ was also faster than both $\times1.5$ ($t(361)=-5.23$, $p<.001$, $d=0.55$) and $\times2.5$ ($t(361)=-4.06$, $p=.009$, $d=0.43$).

\subsection{Subjective Evaluation}
From the preference ranking (Fig. \ref{fig:result3}(B)), the bimanual method was the most preferred technique. Among the unimanual techniques, Push-Pull-Depth scaling was the most preferred. All results showed no significant effects on NASA-TLX factors and subjective questionnaires (Fig. \ref{fig:result3}(C, D)).

\subsection{User Feedback}
\subsubsection{Intuitiveness of Scaling Gestures}
The PTZ scaling gesture was the most mentioned gesture perceived as intuitive ($N=9$). P4 reported being already familiar with the it. On the other hand, depth-based Push-Pull scaling was perceived as less intuitive, requiring participants to anticipate the direction of hand movement before scaling ($N=8$). 

\subsubsection{Perceived Range of Scaling}
Participants noted that bimanual scaling enabled the widest scaling range ($N=5$). Regarding PTZ gestures, incorporating both span and depth (PTZ-Area and PTZ-Angle) was perceived to afford a broader scaling range ($N=2$) than PTZ-Span, which was criticized for its limited range ($N=3$). P5 described that coarse scaling was performed using span and refinement with depth for the PTZ-Angle. In the Push-Pull scaling, participants reported hand tracking loss when their hand was too close to the HMD, causing scaling errors ($N=8$). 

\subsubsection{Relationship between Scaling Accuracy and Mode-out Timing}
Participants noted that the bimanual allowed accurate mode-out at the intended timing, resulting in precise scaling ($N=5$). Specifically, P19 preferred techniques that could accurately reach the target scale through precise mode-out timing. In contrast, participants commented that it was difficult to mode-out at the intended timing with PTZ gestures ($N=9$). 

\subsubsection{Physical Fatigue}
While the baseline of bimanual scaling was the most preferred technique, it was most cited as a physically fatiguing interaction ($N=5$). P4 explained that the method may not be practical for daily life, while P10 mentioned that as the target scale gets bigger, their hand movement also gets bigger.

\section{Discussion} 
\subsection {Comparison with Bimanual Scaling: Performance Costs of Transitioning from Bimanual to Unimanual}
While the results indicate that unimanual scaling did not outperform the baseline of bimanual scaling, we emphasize that the purpose of the research is not to beat the bimanual, but to provide a necessary unimanual alternative. The overlap and angular dispersion-based alignment especially suffered from low scaling precision, because unintended scale change occurred when users moved their hand away from the gaze during mode-out. Instead, the pinch-assisted mode-switch of bimanual and Push-Pull scaling enabled an instant mode-out, resulting in accurate scaling. Furthermore, this performance cost of transitioning from bimanual to unimanual techniques was also observed in previous studies \cite{guiard1987asymmetric, stellmach2012investigating, surale2019experimental}. In this sense, our contribution lies in accommodating users with limited hand availability by demonstrating that Align-to-Scale is learnable and enables intuitive scaling, while simultaneously cautioning designers to account for this inherent performance gap when developing unimanual alternatives.

\subsection{Comparisons among Unimanual Techniques}
\subsubsection{PTZ scaling}
Among the unimanual techniques, PTZ-Area showed the most robust mode-in performance, followed by PTZ-Angle and -Span.
Compared to the conventional angular dispersion cue, the overlap-based alignment may capture users’ intention to interact with a target more quickly and accurately.  
Similar to area cursors, where selection occurs when the interaction area overlaps with the target, it allows faster and easier selection of small or grouped objects~\cite{kabbash1995prince, choi2020bubble}. These benefits, combined with contextual cues from the stereoscopic view area~\cite{wang2022perceptual, lee2023stereoscopic}, may explain its fast and accurate performance. 

However, using the stereoscopic view area as a control parameter was shown to be less robust, due to its poor scaling performance. We attribute this to a fixed threshold of the overlap ratio. When objects became smaller during scale-down, the overlap ratio tends to fall below the threshold more easily before reaching the target scale. In contrast, during $\times 2.5$ scale-up tasks, as the object got bigger, users had to move their hands farther to reduce overlap, causing the scale value to fluctuate during the mode-out. We believe this is why span-based scaling resulted in the lowest scale difference among the control parameters of PTZ, as it was influenced only by span, not depth. 
Considering the feedback, where participants still enjoyed scaling with both span and depth, this conflicting tendency highlights the potential of combining span and depth as a control parameter for scaling.

\subsubsection{Push-Pull Scaling}
The most preferred unimanual scaling method was Push-Pull-Depth-based scaling, which used the pinch-assisted alignment strategy. This was due to its high scaling performance, enabled by precise temporal control over releasing the pinch, afforded by tactile feedback~\cite{huang2016digitspace, waugh2022push}. However, frequent hand-tracking loss near the HMD caused abrupt mode-out during scale-up ($\times 2.5$), increasing scaling errors. This highlights a limitation of depth as a control parameter, where the system’s detectable range and the user’s perceived movement range do not match. Furthermore, participants also mentioned that the Push-pull gesture was not as intuitive as PTZ. Despite the limitations, Push-Pull scaling was still the most preferred unimanual technique, implying that guaranteeing accurate scaling capability is crucial.

\subsection{Design Guidelines for Align-to-Scale}
\subsubsection{Unimanual Techniques for Physically Constrained Conditions}
Our unimanual techniques can be provided as an alternative option to bimanual scaling when users are physically constrained to one-handed interaction. This encompasses not only people with permanent or temporary impairments on upper limbs, but also situational impairments. For instance, unimanual scaling enables multitasking to enhance productivity, such as taking notes with one hand while scaling an object with the other \cite{li2013bezelcursor}. Another common scenario involves users who are encumbered while on the move, often carrying a bag or box with one hand \cite{ng2013impact}. Furthermore, everyday contexts, such as holding a cup or an infant, frequently restrict hand availability, necessitating unimanual interaction \cite{wobbrock2019situationally} and leading users to prefer such techniques \cite{ng2013impact}.

In these scenarios, we recommend choosing a technique driven by task requirements: PTZ scaling when rapid task completion is prioritized over scaling precision, and pinch-assisted Push-Pull gesture for tasks demanding high precision. For instance, when users need to quickly zoom into a map while on the move, PTZ-Area can be used due to the low cognitive load of its intuitive gesture. Similarly, for exploratory tasks like taking a look at distant or minute objects that do not require pixel-wise precision, PTZ scaling can also be used. Instead, for precision-critical scaling such as 3D modeling or technical drawing \cite{lee2024bimanual, jiang2021handpainter}, we recommend the Push-Pull scaling, as it was the only technique to achieve precision comparable to the bimanual baseline.

\subsubsection{Integrating PTZ with Semi-Pinch Quasi-Modes}
To expand quasi-mode interactions specified with the `semi-pinch' gesture~\cite{lee2025facilitating, zhu2023pinchlens}, a pre-pinch state similar to PTZ, can be integrated with Align-to-Scale. For example, in a multi-selection task, users can expand the selection area when the hand aligns with the gaze, and select multiple objects at once when not, effectively adding a scaling dimension to the existing quasi-mode \cite{kim2025pinchcatcher}.

\subsection{Limitations \& Future Directions}
First, our controlled experiment prohibited clutching to isolate mode-switching performance. Because repeated clutching is a common practice when resizing \cite{avery2014pinch}, this constraint may have contributed to higher scaling errors. Future work should investigate clutching-enabled performance to offer more practical insights. Secondly, while we employed visual feedback to indicate the current mode, the feedback method was not tested, as it is beyond the scope of our research. To enhance accessibility, other modalities such as auditory or tactile \cite{jang2024effects} can be considered. Moreover, our design space does not encompass all potential unimanual components. While we focused on gaze-hand alignment as a mode-switch cue, future research could further extend by combining and comparing our Align-to-Scale with other components in Tab.\ref{tab:interaction_comparison}. Lastly, our method may induce arm fatigue, leading to the gorilla arm effect \cite{jang2017modeling}. While we acknowledge that this contrasts with the current Gaze+Pinch design trend in commercial HMDs, which is to reduce arm fatigue, we consider this as a compensation to enable unimanual interactions. 

To improve the scaling accuracy of PTZ, combining different alignment strategies and control parameters could be an option. Future designs might, for instance, pair overlap-based alignment with a span parameter with a PTZ gesture. To make users aware of when the mode-switching occurs, visualizing the stereoscopic view area can also help.

\section{Conclusion}
In this paper, we explore how gaze and hand can be used together to enable unimanual object manipulation, with a focus on the one-handed mode-switching and scaling. We adopted the two most common scaling gestures: PTZ and Push-Pull. As these gestures are not fully compatible with the previous solely hand-based mode-switching to separate scaling and other object manipulation actions, we propose gaze-hand-object alignment-based mode-switching. By aligning the hand, gaze, and the object, users can start scaling, and finish by separating them. The techniques were evaluated to prevent user confusion across different modes while maintaining scaling accuracy. We explain the strengths and weaknesses of each technique and provide design guidelines for unimanual object scaling interactions.


\bibliographystyle{ACM-Reference-Format}
\bibliography{references}

\appendix
\end{document}